\documentclass[twocolumn, prb]{revtex4}

\usepackage{graphicx}
\usepackage{dcolumn}
\usepackage{bm}

\begin{document}

\title{ A Renormalization group approach for highly anisotropic 2D Fermion 
systems: application to coupled Hubbard chains }

\author {S. Moukouri}

\affiliation{ Department of Physics and  Michigan Center for 
          Theoretical Physics \\
         University of Michigan, 2477 Randall Laboratory, Ann Arbor MI 48109}

\begin{abstract}
I apply a two-step density-matrix renormalization group method to the 
anisotropic two-dimensional Hubbard model. As a prelude
to this study, I compare the numerical results to the exact one for the
tight-binding  model. I find a ground-state energy which agrees with the 
exact value up to four digits for systems as large as $24 \times 25$.
I then apply the method to the interacting case. I find that for strong
Hubbard interaction, the ground-state is dominated by magnetic correlations.
 These correlations are robust even in the presence of strong frustration.
Interchain pair tunneling is negligible in the singlet and triplet channels
and it is not enhanced by frustration. For weak Hubbard couplings, interchain
non-local singlet pair tunneling is enhanced and magnetic correlations are
strongly reduced. This suggests a possible superconductive ground state. 
\end{abstract}

\maketitle

\section{Introduction}

Quasi-one dimensional organic \cite{reviewOC} and inorganic \cite{reviewIOC}
 materials have been the object
of an important theoretical interest for the last three decades.
The essential features of their phase diagram may be captured by 
the anisotropic Hubbard model (AHM),

\begin{eqnarray}
\nonumber H=-t_{\parallel}\sum_{i,l,\sigma}(c_{i,l,\sigma}^{\dagger}
c_{i+1,l,\sigma}+h.c.)
 +U\sum_{i,l} n_{i,l,\uparrow}n_{i,l,\downarrow} \\
\nonumber + V\sum_{i,l,\sigma}n_{i,l, \sigma}
n_{i+1,l,\sigma} 
-\mu \sum_{i,l,\sigma} n_{i,l,\sigma} \\
 -t_{\perp}\sum_{i,l,\sigma}(c_{i,l,\sigma}^{\dagger}c_{i,l+1,\sigma}+h.c.).  
\label{hamiltonian}
\end{eqnarray}
 
\noindent or a more general Hubbard-like model including longer range Coulomb 
interactions. The indices $i$ and $l$ label the sites and the chains
respectively. 
For these highly anisotropic materials, $t_{\perp} \ll t_{\parallel}$.  
Over the years, the AHM has remained a formidable
challenge to condensed-matter theorists. Some important insights on this 
model or its low energy version, the g-ology model, have been obtained through 
the work of Bourbonnais and Caron \cite{bourbonnais,giamarchi} and others.
 They used a perturbative renormalization group approach to analyze the
crossover from 1D to 2D at low temperatures. More recently, Biermann et al.
\cite{bierman} applied the chain dynamical mean-field approach to study 
the crossover
from Luttinger liquid to Fermi liquid in this model. Despite this 
important progress, crucial information such as  the ground-state
 phase diagram, or most notably, whether the AHM 
 displays superconductivity, are still unknown. 
So far it has remained beyond the reach of numerical methods such as
the exact diagonalization (ED) or the quantum Monte Carlo (QMC) methods.
ED cannot  exceed lattices of about $4 \times 5$. It is likely to
remain so for many years unless there is a breakthrough in quantum
computations. The QMC method is plagued by the minus sign problem
and will not be helpful at low temperatures. The small value of $t_{\perp}$
implies that, in order to see the 2D behavior,  it will be necessary to reach 
lower temperatures than those usually studied for the isotropic 2D Hubbard 
model. Hence, even in the absence of the minus sign problem, in order to
work in this low temperature regime, the QMC algorithm requires special
stabilization schemes which lead to prohibitive cpu time. \cite{white-qmc}   

\section{Two-step DMRG}

I have shown in Ref.~\onlinecite{moukouri-TSDMRG} that this class of anisotropic 
models may  be studied using a two-step density-matrix renormalization
group (TSDMRG) method. The TSDMRG method is a perturbative approach
in which the standard 1D DMRG is applied twice.
In the first step, the usual 1D DMRG method \cite{white} is applied
to find a set of low
lying eigenvalues $\epsilon_n$ and eigenfunctions $|\phi_n \rangle$ of a
single chain. In the second step, the  2D Hamiltonian is then projected
onto the basis constructed from the tensor product of the $|\phi_n \rangle$'s.
This projection yields an effective one-dimensional Hamiltonian for
the 2D lattice,

\begin{eqnarray}
 \tilde{H} \approx \sum_{[n]} E_{\parallel [n]} |\Phi_{\parallel [n]}
\rangle \langle\Phi_{\parallel [n]}| -
 t_{\perp}\sum_{i,l,\sigma}(\tilde{c}_{i,l,\sigma}^{\dagger}
\tilde{c}_{i,l+1,\sigma}+h.c.)
\end{eqnarray}

\noindent where  $E_{\parallel [n]}$ is the sum of eigenvalues of the
different chains, $E_{\parallel[n]}=\sum_l{\epsilon_{n_l}}$;
$|\Phi_{\parallel [n]}\rangle$ are the corresponding eigenstates,
$|\Phi_{\parallel [n]}\rangle =  |\phi_{n_1}\rangle  |\phi_{n_2}\rangle ...
|\phi_{n_L} \rangle$; $\tilde{c}_{i,l,\sigma}^{\dagger}$, 
$\tilde{c}_{i,l,\sigma}$, and $\tilde{n}_{i,l,\sigma}$
are the  renormalized matrix elements in the single chain basis. They are 
given by

\begin{eqnarray}
(\tilde{c}_{i,l,\sigma}^{\dagger})^{n_l,m_l}=(-1)^{n_i}\langle \phi_{n_l}|
{c}_{i,l,\sigma}^{\dagger}
|\phi_{m_l}\rangle, \\
(\tilde{c}_{i,l,\sigma})^{n_l,m_l}=(-1)^{n_i}\langle \phi_{n_l}|{c}_{i,l,\sigma} |\phi_{m_l}\rangle,\\ 
(\tilde{n}_{i,l,\sigma})^{n_l,m_l}=\langle \phi_{n_l}|{n}_{i,l,\sigma} |\phi_{m_l}\rangle, 
\end{eqnarray}

\noindent where $n_i$ represents the total number of fermions from sites $1$ to
$i-1$. For each chain, operators for all the sites are stored in a
single matrix

\begin{eqnarray}
\label{bm1}
\tilde{c}_{l,\sigma}^{\dagger}=(\tilde{c}_{1,l,\sigma}^{\dagger},...,
\tilde{c}_{L,l,\sigma}^{\dagger}),\\
\label{bm2}
\tilde{c}_{l,\sigma}=(\tilde{c}_{1,l,\sigma},...,\tilde{c}_{L,l,\sigma}),\\
\tilde{n}_{l,\sigma}=(\tilde{n}_{1,l,\sigma},...,\tilde{n}_{L,l,\sigma}).
\end{eqnarray}

\noindent Since the in-chain degrees of freedom have been integrated out,
the interchain couplings are between the block matrix operators in
Eq.(~\ref{bm1},~\ref{bm2}) which depend only on the chain index $l$.
In this matrix notation, the effective Hamiltonian is
one-dimensional and it is also studied by the DMRG method. The only
difference compared to  a normal 1D situation is that the local operators are now
$ms_2 \times ms_2$ matrices, where $ms_2$ is the number of states kept 
during the second step. 

The two-step method has previously been applied to anisotropic
two-dimensional Heisenberg models.\cite{moukouri-TSDMRG} 
In Ref.~\onlinecite{moukouri-TSDMRG2},
it was applied to the $t-J$ model but due to the absence  of an exact
result in certain limits, it was tested against ED results on small 
ladders only.
 A systematic analysis of its performance on a fermionic model on 2D lattices 
of various size has not been done. In this paper, as a prelude to the study of 
the AHM, I will apply the TSDMRG to the 
anisotropic tight-binding model on a 2D lattice, i.e., 
model (\ref{hamiltonian}) with $U=V=0$. 
I perform a comparison with the exact result of the tight-binding model. 
I was able to obtain agreement for the ground-state energies on the order of 
$10^{-4}$ for lattices of up to $24 \times 25$. I then discuss how these
calculations may be extended to the interacting case, before presenting 
the $U \neq 0$ results. 

\section{Warm up: the tight-binding model}

The tight-binding Hamiltonian is diagonal in the momentum space, 
the single particle energies are,

\begin{eqnarray} 
\epsilon_k=-2t_{\parallel} cos k_x-2t_{\perp}cos k_y-\mu,
\label{spectrum} 
\end{eqnarray}

\noindent with $k=(k_x,k_y)$, $k_x=n_x\pi/(L_x+1)$ and $k_y=n_y\pi/(L_y+1)$ for
open boundary conditions (OBC); $L_x$, $L_y$ are respectively the linear
dimensions of the lattice in the parallel and transverse directions. 
The ground-state
energy of an $N$ electron system is obtained by filling the lowest states 
up to the Fermi level,
$E_{[0]}(N)=\sum_{k < k_F}  \epsilon_k$. However in real space, this problem
is not trivial and it constitutes, for any real space method
such as the TSDMRG, a test having the same level of
difficulty as the case with $ U \neq 0$. This is because the term involving 
$U$ is diagonal in real space and the challenge of diagonalizing 
the AHM arises from the hopping term.

I will study the tight-binding model at quarter filling, $N/L_xL_y=1/2$, the
nominal density of the organic conductors known as the Bechgaard salts.
 Systems of up to $L_x\times L_y=L \times (L+1)=24 \times 25$ will be studied.
During the first step, I keep enough states ($ms_1$ is a few hundred)
so that the truncation error $\rho_1$ is less than $10^{-6}$. I
target the lowest state in each charge-spin sectors $N_x \pm 2,~ N_x \pm 1,~ N_x$
and $S_z \pm 1,~ S_z \pm 2$, $N_x$ is the number of electrons within the
chain. It is fixed such that $N_x/L_x=1/2$. There is a total of $22$ charge-spin states 
targeted at each iteration.

\begin{figure}
\includegraphics[width=3. in, height=2. in]{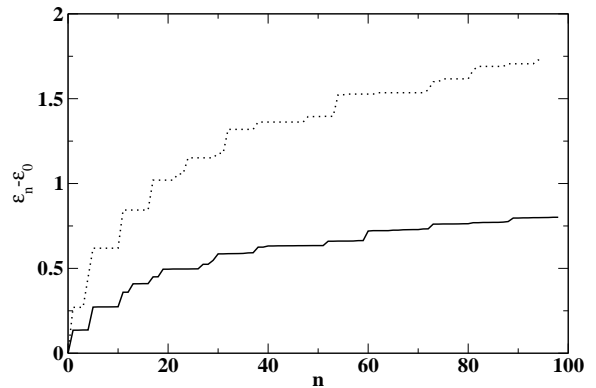}
\caption{Low-lying states of the 1D tight-binding model (full line) and
of the 1D Heisenberg spin chain (dotted line) for $L=16$ and $ms_2=96$. }
\vspace{0.5cm}
\label{density}
\end{figure}

For the tight-binding model, the chains remain disconnected if 
$t_{\perp} < \epsilon_0(N_x+1)-\epsilon_0(N_x)$ or $t_{\perp} < \epsilon_0(N_x)
-\epsilon_0(N_x-1)$, where $N_x$ is the number of electons on single chain.
 In order to observe transverse motion, it is necessary that at least 
$t_{\perp} \agt \epsilon_0(N_x+1)-\epsilon_0(N_x)$ and $t_{\perp} \agt \epsilon_0(N_x) -\epsilon_0(N_x-1)$. These two conditions are satisfied only if $\mu$ is 
appropietly chosen.
 The values listed in Table (\ref{param}) corresponds to 
$\mu= (\epsilon_0(N_x+1)-\epsilon_0(N_x-1))/2$. This treshold varies with $L$. I give
in Table (\ref{param}) the values of $t_{\perp}$ chosen for different
lattice sizes. In principle, for the TSDMRG to be accurate, it is necessary 
that $\Delta \epsilon=\epsilon_{n_c}-\epsilon_{0}$, where $\epsilon_{n_c}$ is 
the cut-off, be such that $\Delta \epsilon/t_{\perp} \gg 1$.
But in practice, I find that I can achieve accuracy up to the fourth 
digit even if $\Delta \epsilon/t_{\perp} \approx 5$ using the finite system
method. Five sweeps were necessary to reach convergence. Note that this 
conclusion is somewhat different from my earlier estimate of 
$\Delta \epsilon/t_{\perp} \approx 10$ for spin systems. \cite{moukouri-TSDMRG2} This 
is because in Ref.~\onlinecite{moukouri-TSDMRG2}, I used the infinite system method 
during the second step.  

The ultimate success of the TSDMRG depends on the density of the
low-lying states in the 1D model. For fixed $ms_2$ and $L$, it is, for 
instance, easier to reach larger $\Delta \epsilon/J_{\perp}$ in the anisotropic spin
one-half Heisenberg model, studied in  Ref.~\onlinecite{moukouri-TSDMRG},
than  $\Delta \epsilon/t_{\perp}$ for the tight-binding model as shown 
in Fig.\ref{density}. For $L=16$, $ms_2=96$, and $J_{\perp}=t_{\perp}=0.15$,
I find that $\Delta \epsilon/J_{\perp} \approx 10$, 
while $\Delta \epsilon/t_{\perp} \approx 5$. Hence, the TSDMRG method will be more 
accurate for a spin model than for the tight-binding model. Using the infinite
system method during the second step on the anisotropic Heisenberg model
with $J_{\perp}=0.1$, I can now reach an agreement of about
$10^{-6}$ with the stochastic QMC method.

\begin{table}
\begin{ruledtabular}
\begin{tabular}{cccc}
  & $8 \times 9$ & $16 \times 17$ & $24 \times 25$ \\
\hline
 $t_{\perp}$ &0.28 & 0.15& 0.1  \\
 $\mu$ & -1.2660 & -1.3411& -1.3657 \\
$\Delta \epsilon/t_{\perp}$ & 6.42 & 5.40 & 5.78 \\ 
\end{tabular}
\end{ruledtabular}
\caption{Transverse hopping and chemical potential used in the 
simulations for different lattice sizes}
\label{param}
\end{table}

Two possible sources of error can contribute to reduce the accuracy
in the TSDMRG with respect to the conventional DMRG. They are  the
truncation of the superblock from $4 \times ms_1$ states to only $ms_2$
states and the use of three blocks instead of four during the second
step. In Table (\ref{eg1}) I analyze the impact of the reduction of
the number of states to $ms_2$ for three-leg ladders. The choice of 
three-leg ladders is motivated by the fact that at this point, the
TSDMRG is equivalent to the exact diagonalization of three reduced
superblocks. It can be seen that as far as 
$t_{\perp} \agt \epsilon_0(N_x+1)-\epsilon_0(N_x)$ and
 $t_{\perp} \agt \epsilon_0(N_x) -\epsilon_0(N_x-1)$, 
  the TSDMRG at this point
is as accurate as the 1D DMRG. Note that
the accuracy remains nearly the same irrespective of $L$ as far as the 
ratio $\Delta \epsilon/t_{\perp}$ remains nearly constant.
Since $\Delta \epsilon$ decreases when $L$ increases, $t_{\perp}$ must be
decreased in order to keep the same level of accuracy for fixed
$ms_2$. In principle, following this prescription, much larger
systems may be studied. $\Delta \epsilon/t_{\perp}$ does not have to be very 
large, in this case it is about $5$, to obtain very good agreement with
the  exact result.

\begin{table}
\begin{ruledtabular}
\begin{tabular}{cccc}
 $ms_2$ & $8 \times 3$ & $16 \times 3$ & $24 \times 3$ \\
\hline
 $64$ & -0.241524 & -0.211929 & 0.204040 \\
 Exact  & -0.241524&  -0.211931 & 0.204049 \\
\end{tabular}
\end{ruledtabular}
\caption{Ground-state energies of three-leg ladders.}
\label{eg1}
\end{table}

The second source of error is related to the fact that   the
effective single site during the second step is now a chain
having $ms_2$ states, I am thus   forced to use three blocks instead 
of four to reduce the computational burden. In Table (\ref{eg2}),
it can be seen that this results in a reduction in accuracy of 
about two orders of magnitude with respect to those of three 
leg-ladders. These results are nevertheless very good given the 
relatively modest computer power involved. All calculations were
done on a workstation.

\begin{table}
\begin{ruledtabular}
\begin{tabular}{cccc}
 $ms_2$ & $8 \times 9$ & $16 \times 17$ & $24 \times 25$ \\
\hline
 $64$ & -0.24761 & -0.21401 & 0.20504 \\
 $100$ & -0.24819 & -0.21414 & 0.20509\\
$120$& -0.24832 & -0.21419 & \\
 Exact  & -0.24857& -0.21432 & 0.20519\\
\end{tabular}
\end{ruledtabular}
\caption{Ground-state energies for different lattice sizes; 
a  single  state was targeted in the second step.}
\label{eg2}
\end{table}

The DMRG is less accurate when three blocks are used instead of four.
This can be understood  by applying the following view on the formation
of the reduced density matrix. The construction of the reduced density matrix
may be regarded as a linear mapping $u_\Psi:~ {\bf F^*} 
\rightarrow {\bf E}$, where ${\bf E}$ is the system, ${\bf F}$ is the 
environment and, ${\bf F^*}$ is the dual space of ${\bf F}$. 
Using the decomposition
of the superblock wave function $\Psi_{[0]}= \sum_i \phi_i^L \otimes \phi_i^R$,
with $\phi_i^L \in {\bf E}$ and $\Phi_i^R \in {\bf F}$, 
for any $\phi^* \in F^*$, 

\begin{eqnarray}
u_\Psi(\phi^*)=\sum_{i=1} \langle\phi^*|\phi_i^R\rangle\phi_i^L.
\end{eqnarray}

\noindent Let $|k\rangle,~k=1,...dim{\bf E}$ and $|l\rangle,~l=1,...dim{\bf F}$
be the basis of ${\bf E}$ and ${\bf F}$ respectively. Then, $|l\rangle$ has 
a dual basis $\langle l^*|$ such that $\langle l^*|l\rangle=\delta_{l,l^*}$.
The matrix elements of $u_\Psi$ in this basis are just the coordinates of
the superblock wave function $\Phi_{[0]_{k,l}}$. The rank $r$ of this mapping,
which is also the rank of the reduced density matrix is
always smaller or equal to the smallest dimension of ${\bf E}$ or 
${\bf F}$, $r < Min(dim {\bf E}, dim {\bf F})$. Hence, if $ms_2$ states are 
kept in the two external blocks, the number of non-zero eigenvalues of $\rho$ 
cannot be larger than $ms_2$. Consequently, some states which have 
non-zero eigenvalues in the normal four block configuration will be missing.
 A possible cure to this problem
is to target additional low-lying states above $\Psi_{[0]}(N)$. The weight of
these states in $\rho$ must be small so that their role is simply to add
the missing states not to be described accurately themselves. A larger weight 
on these additional states would  lead to the reduction of the accuracy for a 
fixed $ms_2$. In table (\ref{eg3}), I show the improved energies when, 
besides the ground state, I target the lowest states of
the spin sectors $S_z=-1$ and $S_z=+1$ with $N$ electrons. The weights were 
respectively
$0.995$, $0.0025$, and $0.0025$ for the three states.  This lowers $E_{[0]}(N)$
in all cases, but the gain does not appear to be spectacular. But I do
not know whether this is due to my choice of perturbation of $\rho$ 
or whether even the algorithm with four blocks would not yield better
$E_{[0]}(N)$. If the lowest sectors with $N+1$ and $N-1$ electrons which
have $S_z=\pm 0.5$ are projected instead, I find that the results are 
similar to those with $S_z=\pm 1$ sectors, there are possibly many ways
to add the missing states. A more systematic approach to this problem has 
recently been suggested. \cite{white2} It is based on using a local
perturbation to build a correction to the density matrix from the site at the 
edge of the system. Here, such a perturbation would be $\Delta \rho= \alpha
c_l^{\dagger} \rho c_l$, where $\alpha$ is a constant, 
$\alpha \approx 10^{-3}-10^{-2}$, and $c_l^{\dagger},~c_l$ are the creation
and annihilation operators of the chain at the edge of the system.
This type of perturbation resulted in an accuracy gain of more than
an order of magnitude in the case of a spin chain. \cite{white2} The three
block method was found to be on par with the four block method. It will
be interesting to see in a future study how this type of local perturbation
performs within the TSDMRG.    

\begin{table}
\begin{ruledtabular}
\begin{tabular}{ccc}
 $ms_2$ & $8 \times 9$ & $16 \times 17$  \\
\hline
 $64$ & -0.24803 & -0.21401  \\
 $100$ & -0.24828 & -0.21417 \\
 Exact  & -0.24857&  -0.21432\\
\end{tabular}
\end{ruledtabular}
\caption{Ground-state energies for different lattice sizes; 
three states were targeted in the second step: the ground state
itself and the lowest states of $S_z=0$ and $S_z=1$ sectors.}
\label{eg3}
\end{table}

\begin{figure}
\includegraphics[width=3. in, height=2. in]{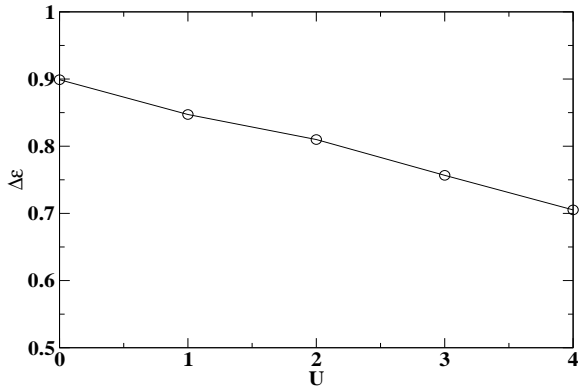}
\caption{Width $\Delta \epsilon$ for the low-lying states of the 1D Hubbard chain
as function of $U$ for  $L=16$ and $ms_2=128$. }
\vspace{0.5cm}
\label{rolu}
\end{figure}

To conclude this section, as a first step to the investigation of interacting 
electron 
models, I have shown that the TSDMRG can successfuly be applied to the  
tight-binding model. The agreement with the exact result is very good and can 
be improved since the computational power involved in this study
was modest. The extension to the AHM with $U \neq 0$
is straightforward. There is no additional change in the algorithm 
since the term involving $U$ is local and thus treated during the
1D part of the TSDMRG. The role of $U$ is to reduce $\Delta \epsilon$ as 
shown in Fig.\ref{rolu}. For fixed $L$ and $ms_2$, $\Delta \epsilon$ decreases
linearly with increasing $U$. For $L=16$ and $ms_2=128$, I anticipate that
for $U \alt 3$ the interacting system results will be on the same level or
better than those of the non-interacting case with $ms_2=100$ for the same 
value of $L$.

\section{Ground-state properties of coupled Hubbard chains}

I now proceed to the study of $U \neq 0$. One of the main motivations
for such a study is the possibility to gain insight into the mechanism
of superconductivity in quasi 1D systems. 
The mechanism of superconductivity in the quasi 1D 
organic materials Bechgaard and Fabre salts, is still an open issue. 
\cite{dupuis} Since these materials are 1D above a crossover temperature 
$T_x \approx t_{\perp}/\pi$, it is broadly accepted that the starting point 
for the the understanding of their low $T$ behavior should be pure 1D physics.
The occurence of the low $T$ ordered phases is driven by the interchain
hopping $t_{\perp}$. Two main hypotheses have been suggested  
concerning superconductivity. The first hypothesis (see a recent review in
Ref.~\onlinecite{dupuis}) relies on a more 
conventional physics: $t_{\perp}$ drives the system to a 2D electron gas 
which is an anisotropic Fermi liquid which becomes superconductive through
 a conventional BCS mechanism. However, it has been argued \cite{emery} 
that given the smallness of $t_{\perp}$, 
the resulting electron-phonon coupling would not be enough to account for the 
observed $T_c$. The second hypothesis, which has gained strength over the years
 given the absence of a clear phonon signature, is that the pairing mechanism
originates from an exchange of spin fluctuation. \cite{emery}

Interest in this issue  was recently revived by the NMR Knight shift 
experimental finding that the symmetry of the Cooper pairs is 
triplet \cite{lee} in $(TMTSF)_2(PF)_6$. No shift was found in the magnetic
susceptibility at the
transition for measurement made under a magnetic field of about 
1.4 {\it Tesla}. A triplet pairing scenario was subsequently supported
by the persistence of superconductivity under fields far exceeding the
Pauli breaking-pair limit \cite{lee2}. However there is no simple explanation of
this scenario. Triplet pairing would be unfavorable in
 a BCS like scenario for which  a singlet s-wave is most likely. Triplet
pairing is also less likely in the spin fluctuation mechanism for which  
a singlet d-wave is predicted by anlytical RG\cite{dupuis} or by 
perturbative approaches \cite{kuroki}. 
It has be argued that these difficulties 
in both mechanisms can be circumvented. In the BCS case, the association
of AFM fluctuations with an open Fermi surface to the electron-phonon
mechanism may lead to a triplet pairing \cite{kohmoto}. In the spin
fluctuation case, the addition of interchain Coulomb interactions may favor 
a triplet f-wave in lieu of the singlet d-wave \cite{dupuis,kuroki}. The more
exotic Fulde-Ferrel-Larkin-Ovchinnikov phase  can also been invoked to
account for the large paramagnetic limit. However,  
the Knight shift  result which was thought to bring a conclusion to 
this long standing issue has only revived the 
old controversy.  The conclusion of this experiment itself
 has been recently challenged. In Ref.~\onlinecite{jerome}, it was pointed out
that the observation of triplet superconductivity claimed in 
Ref.~\onlinecite{lee} could be a spurious effect due to the lack of 
thermalization of the samples. A  recent Knight shift experimenent performed
at lower fields reveals a decrease in the spin susceptibility. This is
consistent with singlet pairing.\cite{shinagawa}

\begin{figure}
\includegraphics[width=3. in, height=2. in]{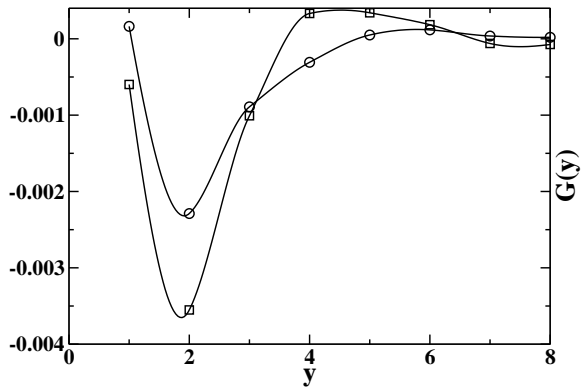}
\caption{Transverse Green's function $G(y)$ for $t_d=0$ (circles),
        $t_d=0.1$ (squares).}
\vspace{0.5cm}
\label{green}
\end{figure}

\begin{figure}
\includegraphics[width=3. in, height=2. in]{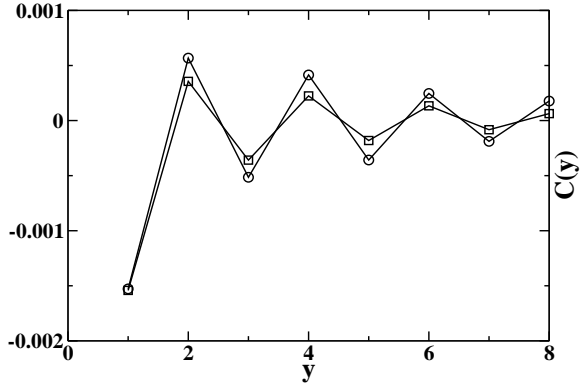}
\caption{Transverse spin-spin correlation $C(y)$ for $t_d=0$ (circles),
        $t_d=0.1$ (squares).}
\vspace{0.5cm}
\label{magn}
\end{figure}

\noindent 
The 1D interacting electron gas is now fairly well understood. 
\cite{bourbonnais} There is no phase with long range order. There
are essentially four regions in the phase diagram, characterized by the
dominant correlations i.e., SDW, charge density wave
(CDW), singlet superconductivity (SS) and triplet superconductivity (TS).
The essential question is whether the interchain hopping will simply
freeze the dominant 1D fluctuation into long-range order (LRO) or 
create new 2D physics. The estimated values of $U$ and  $V$ for
the Bechgaard salts suggest that they are in the SDW region 
in their 1D regime. This suggests that superconductivity in these materials
is a 2D phenomenon. Interchain pair tunneling was suggested
soon after the discovery of superconductivity in an organic 
compound.\cite{jerome-shulz} Emery argued instead that a mechanism similar
to the Kohn-Luttinger mechanism might be responsible for superconductivity
in the organic materials. When $t_{\perp}$ is turned on, pairing
can arise from exchange of short-range SDW fluctuations. The reason is
that the oscillating SDW susceptibility at $Q=(2k_F,k_{\perp})$ would have an
attractive  region if $k_{\perp} \neq 0$. In particular if $k_{\perp}=\pi$ as I 
found, then the interaction would be attractive between particles
in neighboring chains. In this study, I will restrain myself to the study of 
 interchain pair tunneling. 
I was unable to compute correlation functions of pairs in which each
electron belongs to a different chain. The reason is that in the DMRG
method, for the correlation functions to be accurate, at least two
different blocks should be involved. This means that for pair correlation
for which each electron of the pair is on a different chain, at least
four blocks are needed. However, the introduction of four blocks in
the second step of the TSDMRG leads to a prohibitive CPU time.
 
 With the hope of frustrating an SDW ordering which is usually expected, 
I will add an extra terms to model (\ref{hamiltonian}) . These are
the diagonal interchain hopping, 

\begin{eqnarray}
\nonumber H_d= -t_d \sum_{i,l,\sigma}(c_{i,l,\sigma}^{\dagger}c_{i+1,l+1,\sigma}+\\
h.c) +(c_{i+1,l,\sigma}^{\dagger}c_{i,l-1,\sigma} +h.c), 
\label{hamiltonian2}
\end{eqnarray}

\noindent and the next-nearrest neighbor interchain hopping,

\begin{eqnarray}
\nonumber H'_{\perp}= -t'_{\perp} \sum_{i,l,\sigma}(c_{i,l,\sigma}^{\dagger}c_{i,l+2,\sigma}+h.c).
\label{hamiltonian3}
\end{eqnarray}

I will also add the interchain Coulomb interaction,

\begin{eqnarray}
H_{V}= V_{\perp}\sum_{i,l,\sigma}n_{i,l, \sigma}.
n_{i,l+1,\sigma}
\end{eqnarray}

\begin{figure}
\includegraphics[width=3. in, height=2. in]{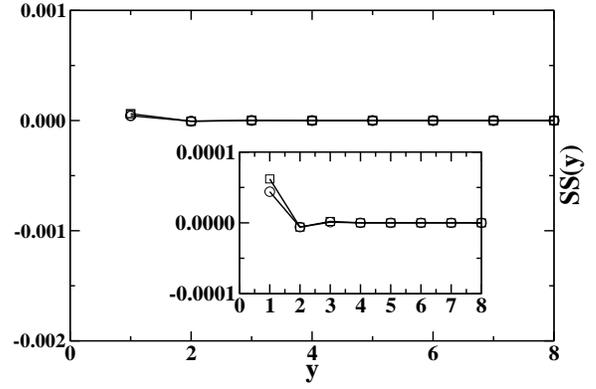}
\caption{Transverse local singlet correlation $SS(y)$ for $t_d=0$ (circles),
        $t_d=0.1$ (squares).}
\vspace{0.5cm}
\label{sups}
\end{figure}

I set $t_{\perp}=0.2$, $ms_1=256$, $ms_2=128$, and
$(L\times(L+1)=16\times17$.
A second set of calculations with $t_{\perp}=0.15$, same values of $ms_1$ and
$ms_2$, and $(L\times(L+1)=24\times25$ lead to the same conclusions. Therefore,
they will not be shown here.
  In order to
analyze the physics induced by the transverse couplings, I compute the following
interchain correlations:
the transverse single-particle Green's function, shown in Fig.\ref{green},
\begin{eqnarray}
G(y)=\langle c_{L/2,L/2+y} c_{L/2,L/2+1}^{\dagger} \rangle,
\end{eqnarray}
\noindent the transverse spin-spin correlation function, shown in 
Fig.\ref{magn},

\begin{eqnarray}
C(y)=\frac{1}{3}\langle {\bf S}_{L/2,L/2+y} {\bf S}_{L/2,L/2+1} \rangle,
\end{eqnarray}
\noindent the transverse local pairs singlet superconductive correlation,
shown in Fig.\ref{sups},

\begin{eqnarray}
SS(y)=\langle \Sigma_{L/2,L/2+y} \Sigma_{L/2,L/2+1}^{\dagger} \rangle,
\end{eqnarray}

\noindent where

\begin{eqnarray}
\Sigma_{i,l}= c_{i,l \uparrow}c_{i,l \downarrow},
\end{eqnarray}
\noindent the transverse triplet superconductive correlation, shown
in Fig.\ref{supp},

\begin{eqnarray}
ST(y)=2 \langle \Theta_{L/2,L/2+y} \Theta_{L/2,L/2+1}^{\dagger} \rangle,
\end{eqnarray}

\noindent where

\begin{eqnarray}
\Theta_{i,l}=\frac{1}{\sqrt{2}} (c_{i,l \uparrow}c_{i+1,l \downarrow} +
c_{i,l \downarrow}c_{i+1,l\uparrow}),
\end{eqnarray}
\noindent and the transverse non-local singlet pair superconductive correlation
function, shown in Fig.\ref{supd},

\begin{eqnarray}
SD(y)=2 \langle \Delta_{L/2,L/2+y} \Delta_{L/2,L/2+1}^{\dagger} \rangle,
\end{eqnarray}

\noindent where

\begin{eqnarray}
\Delta_{i,l}=\frac{1}{\sqrt{2}}( c_{i,l \uparrow}c_{i+1,l \downarrow} - c_{i,l \downarrow}c_{i+1,l\uparrow}).
\end{eqnarray}

\subsection{Strong-coupling regime}

\begin{figure}
\includegraphics[width=3. in, height=2. in]{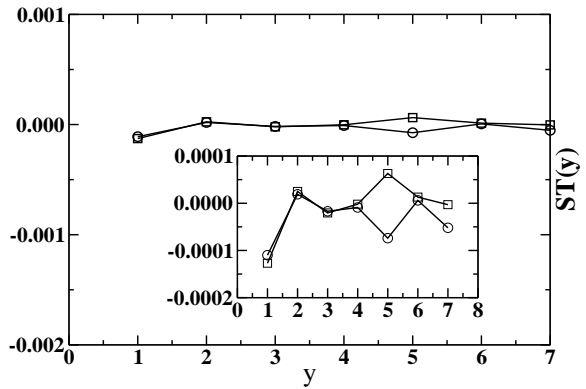}
\caption{Transverse triplet superconductive correlation
$ST(y)$ for $t_d=0$ (circles), $t_d=0.1$ (squares).}
\vspace{0.5cm}
\label{supp}
\end{figure}

\begin{figure}
\includegraphics[width=3. in, height=2. in]{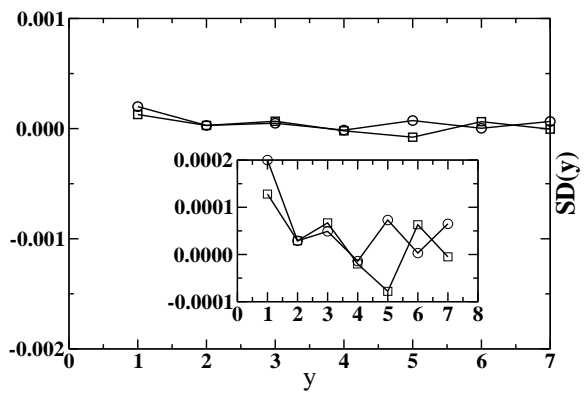}
\caption{Transverse singlet non-local  superconductive correlation $SD(y)$
for $t_d=0$ (circles), $t_d=0.1$ (squares).}
\vspace{0.5cm}
\label{supd}
\end{figure}

 Let us first consider, the regime $U \agt 4$, I choose for
instance $U=4$, $V=0.85$, $\mu=0$, and $t_d=t'_{\perp}=V_{\perp}=0$; besides
single-particle hopping, $t_{\perp}$ also generates two-particle hopping
both in the particle-hole and particle-particle channels. These two-particle
correlation functions are roughly given by the average values
$t_{\perp}^2 \langle  c_{i,l \sigma}^{\dagger}c_{i,l -\sigma}
c_{i,l+j -\sigma}^{\dagger}c_{i,l+j \sigma}\rangle $ and 
$t_{\perp}^2 \langle  c_{i,l \sigma}^{\dagger}c_{i,l -\sigma}^{\dagger} 
c_{i,l+j \sigma}c_{i,l+j -\sigma}\rangle $ for an on-site pair
created at $(i,l)$ and then destroyed at $(i,l+j)$. It is expected
 that the dominant two-particle 
correlation are SDW with $k_{\perp}=\pi$. This is seen in 
Fig.(\ref{magn}-\ref{supd}). The transverse pairing correlations are
all found to be small with respect to $C(y)$. Among the pairing correlations,
$SS(y)$ decays faster then $ST(y)$ and $SD(y)$. These results are consistent
with the view that the role of $t_{\perp}$ is to freeze the dominant 1D
correlations into LRO.

\begin{figure}
\includegraphics[width=2. in, height=2. in]{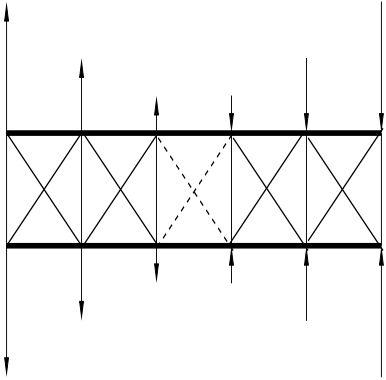}
\caption{sketch of the spin texture (arrows) in two consecutive chains
in an SDW.  The bold horizontal lines represent the chains.
The full diagonal lines 
show bonds for which $t_d$ tends to increase the SDW order. The diagonal
dotted lines show bonds for which $t_d$ frustrates the magnetic order.}
\vspace{0.5cm}
\label{frust}
\end{figure}

When $t_d \neq 0$, it is expected that for a strong enough $t_d$, the
magnetic order will vanish because of the frustration induced by
$t_d$. A simple argument is that
$t_d$ induces an AFM exchange between next-nearest neigbhors on chains
$l$ and $l+1$ which compete with the AFM exchange between nearest
neigbhors. The hope is that there could be a region of the phase
diagram where superconductivity could ultimately win either by pair
tunneling between the chains or by the Emery's mechanism. However,
in Fig.(\ref{magn}-\ref{supd}) it can be seen that, while $t_d$
slightly reduces $C(y)$, the dominant 
correlations are still SDW even for a strong $t_d/t_{\perp}=0.5$. 
$SS(y),~ST(y)$ and $SD(y)$ are barely affected by $t_d$. The fact
that $t_d$ does not strongly affect the SDW order can be understood 
in the light of   recent study
of coupled $t-J$ chains \cite{moukouri-TSDMRG2}. It was shown that the 
frustration strongly suppresses magnetic LRO only close to half-filling. 
For large dopings, two 
neighboring spins in a chain do not always points to opposite direction
as the consequence, $t_d$ does not necessarily frustrate the magnetic
order. This is illustrated in a simple sketch in Fig.(\ref{frust}). 
$t_d$ could even enhance it as seen in the study 
of $t-J$ chains. In Fig.\ref{green}, it can be seen that $t_d$ enhances
$G(y)$. This enhancement, together with the decrease of $C(y)$, suggests a
possible widening of an eventual Fermi liquid region at finite T above the
ordered phase. When $t_{\perp} \neq 0$, I also found (not shown) that 
magnetic correlation are not effectively suppressed even when 
$t'_{\perp}=t_{\perp}/2$. For this value, it would be expected that the
ratio of the
effective exchange term generated by $t'_{\perp}$ to that
generated by $t_{\perp}$ is about one quarter. 
In the frustrated $J_1-J_2$ spin chain, a spin
gap opens around this ratio. This simple picture does not seem to work
here.

\subsection{Weak-coupling regime}

\begin{figure}
\includegraphics[width=3. in, height=2. in]{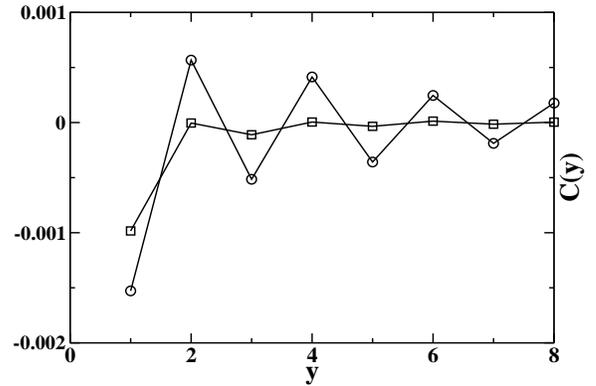}
\caption{Transverse spin-spin correlation $C(y)$ for $U=4$ (circles),
        $U=2$ and $V_{\perp}=0.4$ (squares).}
\vspace{0.5cm}
\label{magn2}
\end{figure}

I now turn in to the regime where $U \alt 4$. I set  $U=2$, $V=0$, 
$\mu=-0.9271$, $t_{\perp}=0.2$, $t_d=0$, and $V_{\perp}=0.4$, 
where $V_{\perp}$ is the interchain Coulomb interaction
between nearest neighbors. It can be seen in Fig.\ref{magn2} that $C(y)$ is now
strongly reduced with respect to its strong coupling values. It is
already within our numerical error for the next-nearest neighbor in
the transverse direction. This is an indication that the ground state
is probably not an SDW. It is to be noted that this occurs even in 
the absence of $t_d$ or $t'_{\perp}$. This seems to be at variance with
the RG analysis which requires $t'_{\perp}$ to destroy the magnetic order.
A possible explanation of this is that at half-filling the perfect nesting
occurs  at the wave vector $Q=(\pi,\pi)$ for the spectrum of 
equation (\ref{spectrum}). Away from half-filling the nesting is no longer 
perfect this leads to the reduction of  magnetic correlations. 
The first correction to the nesting is an effective frustration term which is 
roughly $t_{\perp}^2 cos 2k_{\perp}$. This expression is identical to a term
that could be generated by an explicit frustration $t'_{\perp}=t_{\perp}^2$. 
The discrepancy between the TSDMRG and the RG results could be that this
nesting deviation is underevaluated in the RG analysis. This mechanism cannot
be invoked in the strong coupling regime where band effects are small.

\begin{figure}
\includegraphics[width=3. in, height=2. in]{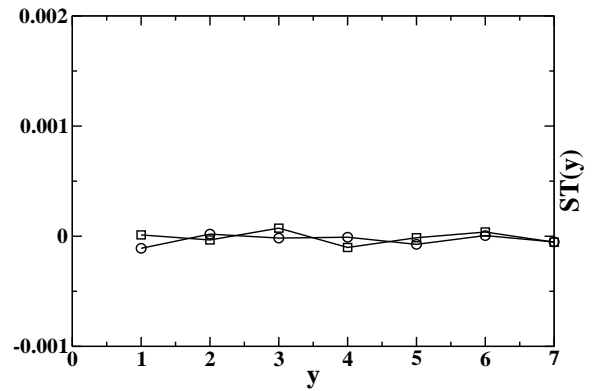}
\caption{Transverse triplet superconductive correlation
$ST(y)$ for $U=4$ (circles), $U=2$ and $V_{\perp}=0.4$ (squares).}
\vspace{0.5cm}
\label{supp2}
\end{figure}

\begin{figure}
\includegraphics[width=3. in, height=2. in]{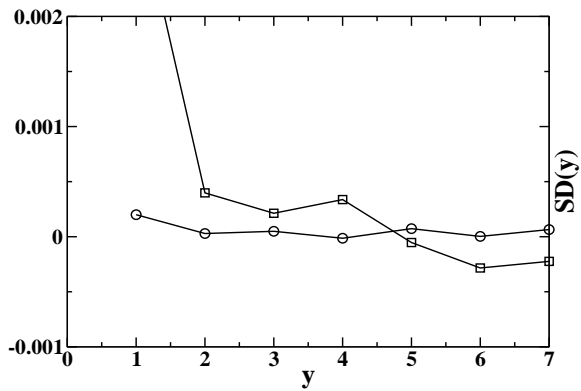}
\caption{Transverse singlet non-local  superconductive correlation $SD(y)$
for $U=4$ (circles), $U=2$ and $V_{\perp}=0.4$ (squares).}
\vspace{0.5cm}
\label{supd2}
\end{figure}

The suppression of magnetism is concommitant to a strong enhancement
of the singlet pairing correlations as seen in Fig. \ref{supd2}.
Triplet correlations,  shown in Fig. \ref{supp2}, remain very small. However,
while it is clear from the behavior of $C(y)$ that the ground state is
non magnetic. This result strongly suggests that the ground state is
a superconductor in this regime. A finite size analysis is, however, 
necessary to conclude whether this persists to the thermodynamic limit. 
I cannot rule out the possibility of a Fermi liquid ground state, which is 
implied by strong single particle correlations. 

\section{Conclusion}

In this paper, I have presented a TSDMRG study of the competition 
between magnetism and superconductivity in an anisotropic Hubbard
model. I have analyzed the effect of the interchain hopping in the strong
and weak $U$ regimes. In the strong-coupling regime,
the results are consistent with earlier predictions that the
role of $t_{\perp}$ is to freeze the dominant 1D SDW correlations into
a 2D ordered state. But at variance with analytical predictions, this
is only true in the strong $U$ regime. In this regime, I find that even
the introduction of frustration does not disrupt the SDW order which remain
robust up to large values of the frustration. In the weak coupling regime
singlet pair correlations are dominant. The ground state seems to be  
a superconductor. This behavior is somewhat in agreement with experiments
in the Bechgaard or Fabre salts. The phase diagram is dominated by magnetism 
at low pressure (strong U) and by superconductivity at high pressure (weak U). 
Because of experimental relevance, I restricted myself to the competion
between magnetism and superconductivity.  I did not analyze CDW 
 correlations. These are likely to be important given that I applied open 
boundary conditions which are known to generate Friedel oscillations \cite{wsa} that very decay slowly from the boundaries.  They may also genuinely generated
by $V_{\perp}$, leading to a CDW ground state instead of a superconductor.

\begin{acknowledgments}
 I am very grateful to C. Bourbonnais for very helpful exchanges. 
I wish to thank A.M.-S. Tremblay for helpful 
discussions. This work was supported by the NSF Grant No. DMR-0426775.

\end{acknowledgments}

\end{document}